\documentclass{caosp}
\usepackage{graphicx}

\articleNo{123}
\pubyear{2008}
\volume{35}
\volnumber{3}
\firstpage{1}
\received{2007}
\accepted{2007}

\begin{document}

\hauthor{M.J.\,Stift and F.\,Leone}

\title{Paschen is Partially Back}

\author{M.J.\,Stift \inst{1} \and F.\,Leone \inst{2}}

\institute{Institut f{\"u}r Astronomie (IfA), Universit{\"a}t Wien,
           T{\"u}rkenschanzstrasse 17, A-1180 Wien, Austria\\
         \and 
           Universit{\`a} di Catania, Dipartimento di Fisica e
           Astronomia -- Sezione Astrofisica, Via~S.~Sofia~78, I-95123
           Catania, Italy
          }

\date{December 1, 2007}

\maketitle

\begin{abstract}
We present a discussion of the partial Paschen-Back (PB) effect in
magnetic Ap stars. An overview of the theory is illustrated with 
examples of how splittings deviate non-linearly from the simple Zeeman 
picture; normally forbidden ``ghost lines'' appear in strong fields. 
Resulting asymmetric stellar Stokes profiles for a dipolar magnetic 
geometry are shown for the Fe\,{\sc ii} $\lambda\,6149$ line and it is 
established that PB lines may be subject to wavelength shifts. Modelling
of Stokes profiles in the PB regime opens exciting new diagnostics.
\end{abstract}

\section{Introduction}
\label{intr}
The discovery in 1897 by P.\,Zeeman of the splitting of spectral lines in 
magnetic fields has provided astrophysicists with a valuable diagnostic tool.
Following the pioneering work of G.E.\,Hale (1908) on sunspots, Zeeman 
observations have been carried out on all kinds of magnetic structures found 
throughout the solar atmosphere, leading to a deeper understanding of the role 
of magnetic fields in atmospheric dynamics. The later discovery by Friedrich 
Paschen and Ernst Back (1921) of the transition in very strong magnetic fields 
from the anomalous Zeeman effect to the normal Zeeman effect has had much less
impact. Indeed, even in fields of 0.4\,T as encountered in sunspots, very few 
spectral lines of special astrophysical interest are affected by the Paschen-Back 
(PB) effect and so only a handful of atomic lines have been analysed for PB
signatures (see e.g. Engvold {\it et al.}, 1970; Socas Navarro {\it et al.}, 2004).

When H.W.\,Babcock ((1960) discovered a 3.4\,T field in the upper main sequence 
chemically peculiar (Ap) star HD\,215441, it should have been obvious that a lot 
of atomic transitions would find themselves somewhere between the Zeeman and the 
full PB regime (the partial or incomplete PB effect). However, not until much later 
was any thought spent on this problem (Kemic, 1975; Stift, 1977). A major discussion 
of the partial PB effect and a line profile modelling attempt is due to Mathys 
(1990); Landolfi {\it et al.} (2001) had a look at the PB effect on hyperfine 
splitting and predicted how this would affect the interpretation of observations.

In strongly magnetic Ap stars, it turns out that high-resolution spectra of the 
Zeeman doublet of Fe\,{\sc ii} at 6149\,{\AA} cannot be fitted with synthetic 
line profiles calculated in the Zeeman approximation. Mathys (1990) attributed
the observed non-symmetric relative intensities of the two components and their 
shift in wavelength (relative to the magnetic null lines of iron) to the partial 
PB effect: at 2\,T the magnetic splitting attains about 35\% of the distance between 
the nearest fine structure levels of the lower term. We decided to have a close look 
at this particular iron line which is heavily used in magnetic field measurements, 
to model it in detail for realistic stellar atmospheres and various magnetic geometries, 
and to search for other spectral lines affected by the partial PB effect.

\section{Theory and computational tools}
\label{tools}
The vector model of the Zeeman effect assumes that the spin-orbit interaction 
between $L$ and $S$ is stronger than their interaction with the magnetic field
$B$. The total angular momentum is $J = L + S$  and $J$ precesses about the 
magnetic field vector $B$, leading to $2 J + 1$ magnetic sublevels characterised
by the quantum number $M$. When instead the magnetic splitting by far exceeds 
the fine-structure splitting, both $L$ and $S$ first interact with the magnetic 
field, and $J$ is no longer a good quantum number. The partial (incomplete) PB 
effect is situated between these 2 extrema; splitting and relative component 
strengths depend non-linearly on magnetic field strength. The energy values of 
the levels (the eigenvalues) and the corresponding eigenvectors now have to 
calculated by diagonalisation of a set of matrices. See Sect. 3.4 of the 
beautiful monograph by Landi Degl'Innocenti \& Landolfi (2004) for details.

\begin{figure}[thp]
\centerline{\includegraphics[width=5.7cm,clip=]{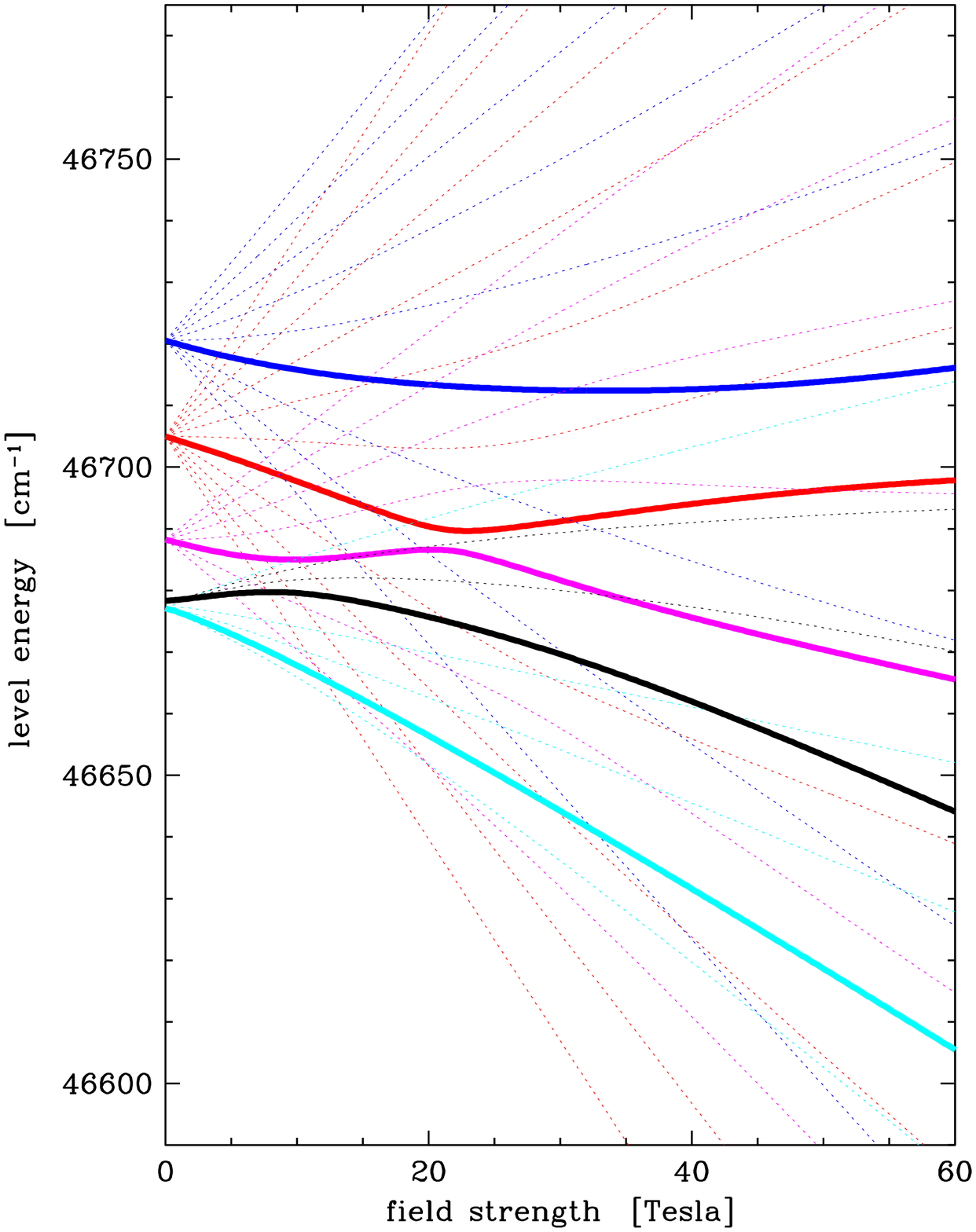} 
            \hspace{0.2cm}
            \includegraphics[width=5.7cm,clip=]{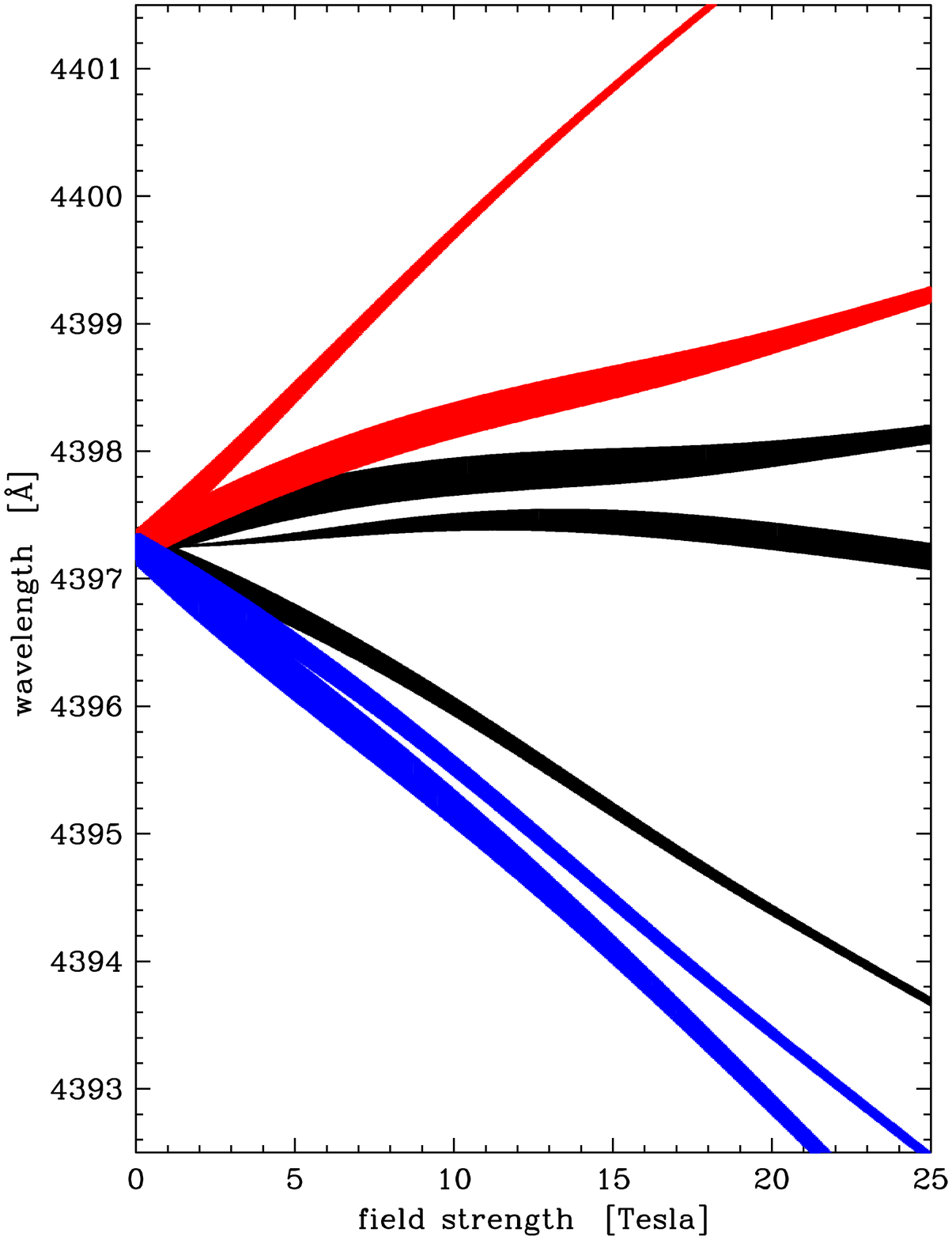}
           }
\caption{{\bf (a)} Avoided crossings in a Cr\,{\sc i} $^5{D}$ term. 
{\bf (b)} Splitting pattern and intensities of the Cr\,{\sc i} line at 
4397.229\,{\AA}.}
\label{f1}
\end{figure}

Thanks to multi-core architectures, detailed stellar polarised line synthesis in 
the partial PB regime has become affordable. Derived from the Stokes code COSSAM 
(Stift 2000), CossamPaschen is the first code ever to allow the correct modelling 
of a multiplet in the partial Paschen-Back regime. Which means realistic stellar 
atmospheres, a sophisticated oblique rotator model (Stift, 1975), component by 
component opacity sampling (CoCoS) (Stift, 2005), and the Zeeman Feautrier 
solution to the polarised RTE (Alecian, Stift 2004). Blending with the other 
transitions in the spectrum (these are treated in the Zeeman approximation) is 
fully taken into account. CossamPaschen requires lots of computer memory because 
splittings and relative subcomponent intensities go non-linearly with field 
strength; so for each point on the stellar surface (depending on field geometry 
and on rotation we need several $10^2 - 10^3$ points) the exact splittings 
and relative intensities for the given local field strength have to be determined 
by diagonalisation of the set of matrices outlined above and stored in appropriate 
data structures. The object-oriented approach (Stift, 1998ab) based on Ada95 
greatly facilitated the necessary modifications.

\begin{figure}[thp]
\centerline{\includegraphics[width=11.5cm,clip=]{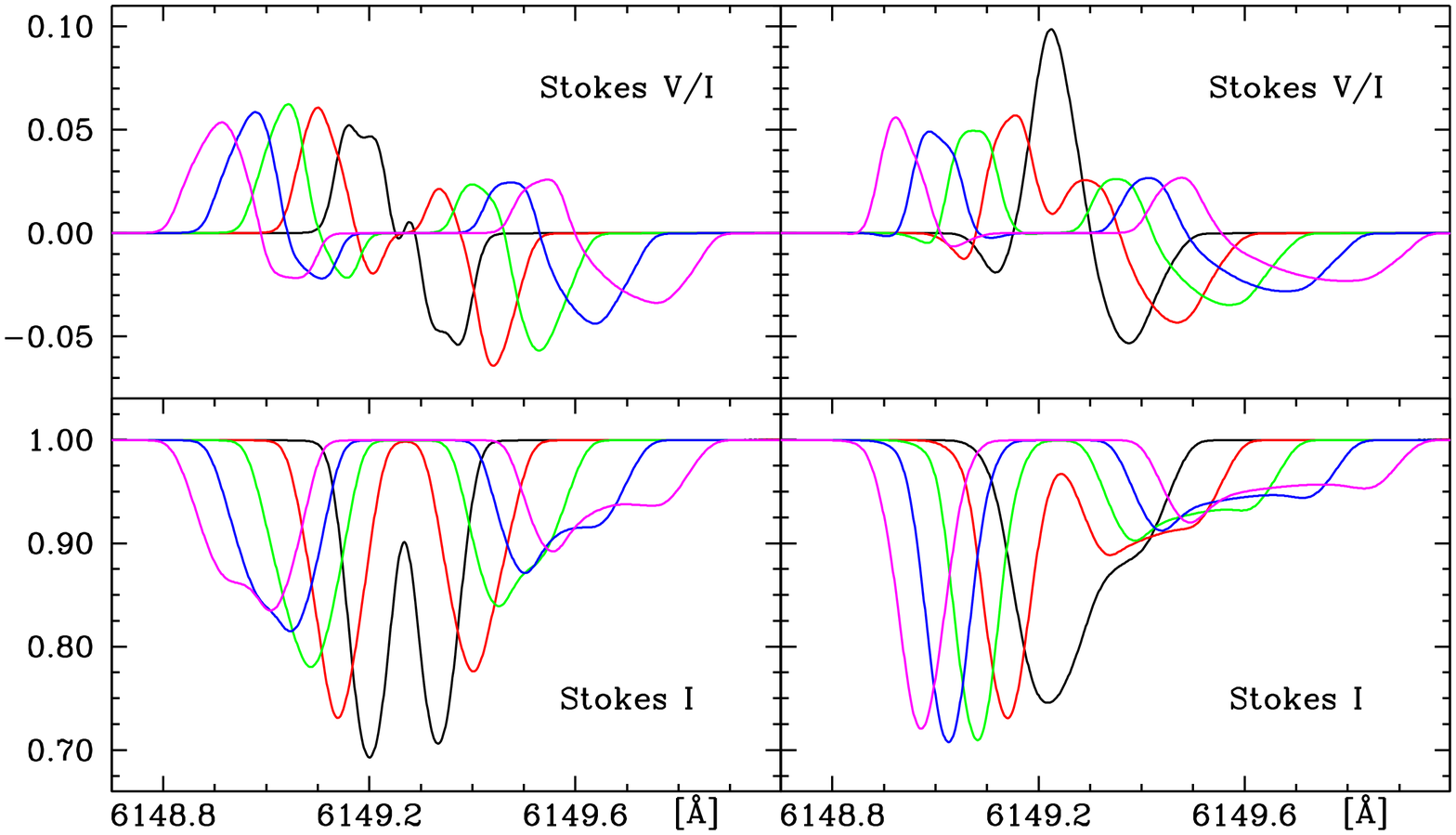} 
           }
\caption{{\bf (Left)} Stokes $I$ and $V$ profiles of the 
Fe\,{\sc i}\,$\lambda\,6149$ line for an oblique rotator model 
(as explained in the text). No rotation, the results pertain to 
field modulus values of $H_s = 0.29, 0.57, 0.86, 1.14,$ and $1.43$\,T.
{\bf (Right)} The same, but for $v\,\sin\,i = 5$\,km/s.}
\label{f2}
\end{figure}

\section{Level splittings}
\label{split}
In the PB regime the splittings of the energy levels in a multiplet change in a 
non-linear and asymmetric way with magnetic field strength. So do the relative 
intensities of the subcomponents. The smallest distance between adjacent levels 
of the ${^4}D$ term of multiplet \,74 of Fe\,{\sc ii} (to which the well-known 
line 6149\,{\AA} belongs) is about 4.0\,cm$^{-1}$ and deviations from linear 
Zeeman splitting are revealed for fields $\geq 0.7$\,T. In the 
Li\,{\sc i}\,$\lambda\,6707$ doublet (just 0.34\,cm$^{-1}$ fine structure 
splitting) asymmetries can occur at much lower field strengths (0.2\,T) but the 
relative intensities are affected even earlier: the two $\pi$-components differ 
by 20\,\% at 0.05\,T, the respective blue and red $\sigma$-components by 10\,\%. 

Sometimes we encounter ``avoided crossings'', so familiar to helioseismologists. 
Sublevels of a spectroscopic term which have the same value of $M$ must not 
cross, whereas the others are free to do so. This is beautifully illustrated
for the $M = -1$ levels of a Cr\,{\sc i} $^5{D}$ term (Fig.\,1a). Such avoided 
crossings are of course reflected in exotic splitting patterns (Fig.\,1b).

\section{The profile of the Fe\,{\sc ii} $\lambda$\,6149 line}
\label{profile}
This famous line, belonging to multiplet 74 of Fe\,{\sc ii}, is at the basis of
most determinations of $H_s$, the field modulus integrated over the visible
hemisphere of a magnetic star. Mathys (1990) has already drawn attention to the
observed aysmmetry of the line profile in the strongly magnetic Ap star HD\,126515
which cannot be explained in the Zeeman approximation (the rotation is almost
negligible), and he invoked the partial Paschen-Back effect as an explanation.
Fig.\,2 shows PB profiles in Stokes $I$ and $V$ for a simple centred dipole 
oblique rotator model and for $v\,\sin\,i = 0$ (left) and 5\,km/s (right panel). 
With $i = 90$\degr and obliquity $\beta = 90$\degr, the star is seen at phase 
$\phi = 0.125$, i.e. the dipole axis is inclined by 45\degr towards the 
line-of-sight. Without rotation, the doublet becomes aysmmetric already at 
$H_s = 0.29$\,T; at $H_s = 1.43$\,T the red component is quite broad and 
shallow, the blue component narrower and deeper. The Stokes $V$ profiles are
not quite as asymmetric.

\begin{figure}[thp]
\centerline{\includegraphics[width=5.85cm,clip=]{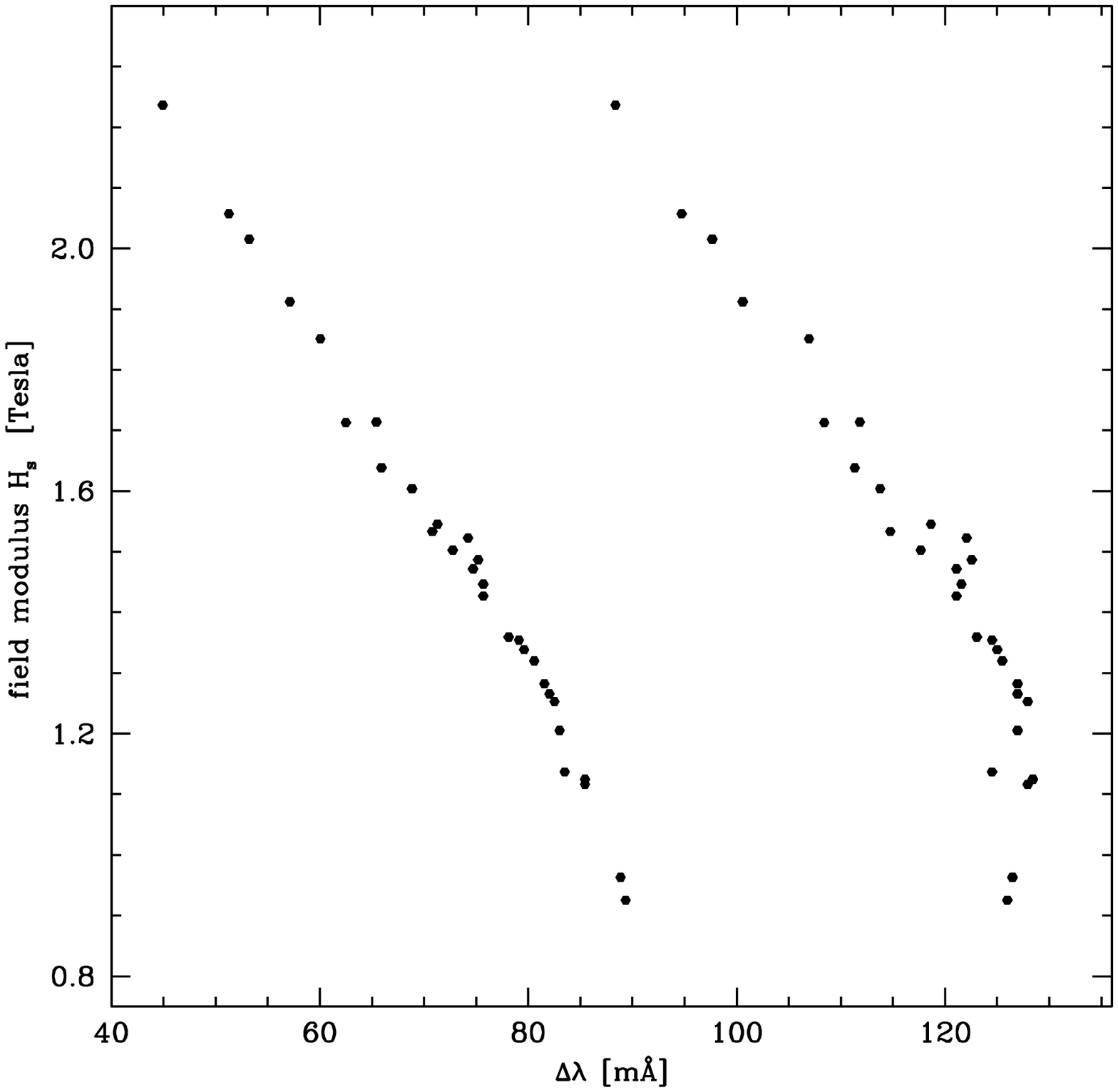} 
            \hspace{0.2cm}
            \includegraphics[width=5.85cm,clip=]{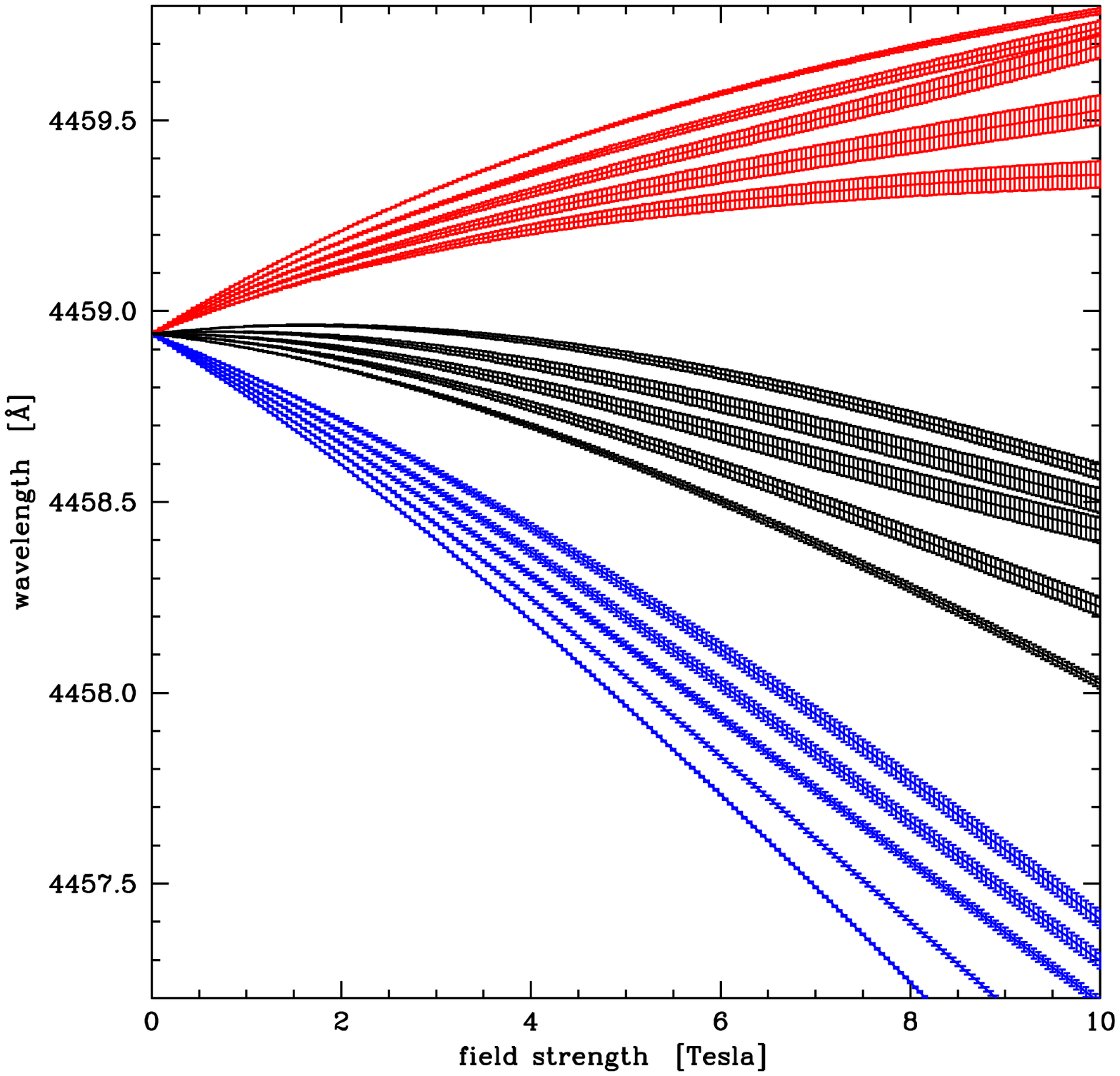}
           }
\caption{{\bf (a)} Wavelength shift of the centre of gravity (in
Stokes $I$) of the combined
Fe\,{\sc ii} $\lambda\,6147$ and $\lambda\,6149$ lines as a function of
mean field modulus $H_s$. The abundance difference between the 2 sets of
points is 1\,dex.
{\bf (b)} Splitting pattern and intensities of the Cr\,{\sc i} ``ghost line'' 
at 4458.92\,{\AA}, belonging to multiplet 127.}
\label{f3}
\end{figure}

With moderate rotation of a mere $v\,\sin\,i = 5$\,km/s, the profiles change 
drastically, and the asymmetry becomes even more pronounced. For the strongest
fields displayed in Fig.\,2 (8.6 - 14.3\,T) we encounter a very narrow blue 
component and an extremely extended red component (with the ratio of the 
respective FWHM approaching a value of 4). These huge asymmetries are reflected 
to a lesser degree in the Stokes $V$ profiles. Detailed modelling of such 
profiles should provide fascinating diagnostic capabilities.

Fortunately, despite these spectacular profile changes due to the PB effect, the 
distance between the respective centres-of-gravity of the blue and the red 
components of $\lambda\,6149$ remain almost exactly the same as in the Zeeman
regime. Measurements of $H_s$ made in the past thus do not have to be revised.

\subsection{Line shifts}
\label{shifts}
Line shifts in Stokes $I$ are a natural consequence of the non-linear splittings 
and the field-dependent relative intensities of the Zeeman subcomponents. Taking
30 different oblique rotator models with $H_s = 0.9 - 2.2$\,T, we find a fairly 
tight relation between $H_s$ and the shift of the centre of gravity of the 
$\lambda\,6147-6149$ blend from its zero field position. Largely insensitive to 
geometry, the relation strongly depends on line strength. Fig.\,3a displays the 
results for 2 abundances differing by 1\,dex. The set of points to the right 
corresponds to the higher abundance and the maximum displacement of about 
125\,m{\AA} occurs near $H_s = 1.2$\,kG. Such shifts in conjunction with the 
strong profile variations could erroneously be interpreted as resulting from 
inhomogeneous element distributions.

\section{Ghost lines}
\label{ghost}
In their monograph, Landi Degl'Innocenti \& Landolfi (2004) discuss in some
detail the appearance in the incomplete Paschen-Back regime of lines which are 
forbidden under the usual selection rules but which originate from J-mixing of 
the various levels of a term. The strength of these lines is a strong function 
of field strength. and so many of these {\em ``ghost lines''} can be of 
comparable strength to the allowed lines when the field exceeds 10-20\,T.

Multiplet 127 of Cr\,{\sc i}, in addition to the 12 allowed lines, gives rise
to 13 ghost lines which at 0.1\,mT are $10 - 35$ orders of magnitude weaker 
than allowed lines. At 10\,T this difference decreases to $0 - 10$ orders 
of magnitude. Fig.\,3b shows the nicely non-linear behaviour of the 
$\lambda\,4458.92$ ghost line. It is not as yet clear whether any ghost lines 
belonging to strong multiplets of overabundant elements could possibly be 
observed -- as suggested by Mathys (1990) -- in Ap stars, where parts of the
surface can exhibit fields of $3 - 5$\,T.

\section{Conclusions and outlook}
\label{out}
A number of atomic lines observed in strongly magnetic Ap stars manifest 
asymmetries and line shifts that cannot be modelled in the Zeeman regime and 
have to be ascribed to the partial Paschen-Back effect. We can show that 
these are not just some exotic few because no less than 846 lines originate 
from a single Cr\,{\sc i} term with fine structure splittings 
$0.25 - 2.77$\,cm$^{-1}$. Whenever magnetic splittings become of comparable 
size, a detailed PB treatment of the whole multiplet containing this term 
is warranted and we have illustrated this with the first ever realistic 
Stokes profile modelling of the Fe\,{\sc ii} $\lambda\,6149$ line. Expecting 
exciting new diagnostic possibilities thanks to such improved modelling, we 
have started further explorations in this fascinating new field.

\acknowledgements
MJS acknowledges support by the {\sf\em Austrian Science Fund (FWF)}, 
project P16003-N05 ``Radiation driven diffusion in magnetic stellar 
atmospheres''. Thanks go to E.\,Landi Degl'Innocenti for his program 
that calculates Paschen-Back patterns and for illuminating comments
and explanations.

\end{document}